\iftrue
\documentclass[aps,prl,twocolumn,
			   groupedaddress,superscriptaddress,
			   amsfonts,amssymb,amsmath,
			   citeautoscript,
			   a4paper]{revtex4-1}
\else
\documentclass[aps,pra,reprint,
			   groupedaddress,superscriptaddress,notitlepage,
			   amsfonts,amssymb,amsmath,
			   citeautoscript,
			   a4paper]{revtex4-1}
\fi

\usepackage[pdftex]{hyperref}
\hypersetup{colorlinks,
			linkcolor={blue!75!black!80!yellow},
			citecolor={blue!75!black!80!yellow},
			urlcolor={blue!75!black!80!yellow},
			pdfstartview=FitH}
\usepackage{graphicx}
\usepackage{filecontents}
\usepackage{mathrsfs}
\usepackage{amsthm}
\usepackage{physics}
\usepackage{xspace}
\usepackage{braket}
\usepackage{xr}
\usepackage{gensymb} 
\usepackage{xcolor,soul}
\usepackage{stmaryrd} 
\usepackage[UKenglish]{babel}
\usepackage{placeins} 
\usepackage{siunitx}
\DeclareSIUnit[number-unit-product=]\percent{\char`\%} 
\usepackage{makecell}
\usepackage{diagbox}
\usepackage{colortbl}

\usepackage{txfonts}  
\usepackage{txfontsb} 

\makeatletter
\def\mathcolor#1#{\@mathcolor{#1}}
\def\@mathcolor#1#2#3{%
  \protect\leavevmode
  \begingroup
    \color#1{#2}#3%
  \endgroup
}
\makeatother

\newcommand*\diff{\mathop{}\mathrm{d}}

\newcommand{\iu}{\mathrm{i}}
\newcommand{\e}{\mathrm{e}} 
\newcommand{\Z}{\mathbb{Z}} 
\newcommand{\bS}{\mathbf{S}}

\newcommand{\para}{\parallel\xspace}

\newcommand{\appropto}{\mathrel{\vcenter{
			\offinterlineskip\halign{\hfil$##$\cr
				\propto\cr\noalign{\kern2pt}\sim\cr\noalign{\kern-2pt}}}}}

\newcommand{\matr}[1]{\boldsymbol{#1}}     
\newcommand{\sigmax}{\sigma_x}
\newcommand{\sigmay}{\sigma_y}

\newcommand{\ie}{i.e.\@\xspace}  
\newcommand{\cf}{cf.\@\xspace}
\newcommand{\eg}{e.g.\@\xspace}
\newcommand{\hdd}{H}

\newcommand{\mtp}{U_{xy}}
\newcommand{\ha}{H_{\mathrm{a}}}
\newcommand{\hb}{H_{\mathrm{b}}}

\makeatletter
\newcommand*{\addFileDependency}[1]{
  \typeout{(#1)}
  \@addtofilelist{#1}
  \IfFileExists{#1}{}{\typeout{No file #1.}}
}
\makeatother

\newcommand*{\myexternaldocument}[1]{%
    \externaldocument{#1}%
    \addFileDependency{#1.tex}%
    \addFileDependency{#1.aux}%
}
\newcommand{\sisec}[1]{Sec.~\ref{#1}}

\usepackage{scalerel}

\usepackage{textcomp} 
\usepackage{xifthen}
\usepackage{etoolbox}
\newboolean{togglecomments}
\newboolean{togglechanges} 

\setboolean{togglecomments}{false}
\setboolean{togglechanges}{true}

\newcommand{\comment}[2]{%
    \ifbool{togglecomments}%
    {\textcolor{blue!70!black}{\small\textsf{%
    \textsuperscript{\textsc{\textsf{\MakeLowercase{#1}}}}%
    [#2]}}} 
    {}}     
\newcommand{\swap}[2]{\ifbool{togglechanges}
    {#2}  
    {\textcolor{red!70!black}{[#1]}\textrightarrow{}\textcolor{green!50!black}{[#2]}}}
\newcommand{\remove}[1]{\ifbool{togglechanges}
    {}    
    {\textcolor{red!70!black}{#1}}}
\newcommand{\inset}[1]{\ifbool{togglechanges}
    {#1}  
    {\textcolor{green!50!black}{#1}}}
\newcommand{\optional}[1]{\ifbool{togglechanges}
    {#1}  
    {\textcolor{gray!30!orange!80!gray}{#1}}}
\newcommand{\citeremind}[1]{%
    [\textcolor{blue!75!black!80!yellow}{
        $\blacksquare$%
           \ifthenelse{\isempty{#1}}
               {}
               {\textsuperscript{\tiny\textsf{#1}}}%
        }]\xspace}
\newcommand{\todo}[1]{
    \textcolor{orange!80!yellow!95!black}{\textbf{[}%
        \ifthenelse{\isempty{#1}}%
        {\text{$\blacksquare$}}%
        {{\small\textsf{#1}}}%
        \textbf{]}}}
        
\newcommand{\mitaffil}{\footnotesize Department of Physics, Massachusetts Institute of Technology, Cambridge, Massachusetts 02139, USA}
\newcommand{\hkuaffil}{\footnotesize Department of Physics, University of Hong Kong, Pokfulam, Hong Kong, China}
\newcommand{\ustaffil}{\footnotesize Department of Physics, Hong Kong University of Science and Technology, Clear Water Bay, Hong Kong, China}
\newcommand{\berkeleyaffil}{\footnotesize Department of Physics, University of California, Berkeley, California 94720, USA}

\myexternaldocument{SM}

\begin{document}



\title{Non-Abelian nonsymmorphic chiral symmetries}



\author{Yi~Yang}
\email{yiyg@hku.hk}
\affiliation{\mitaffil}
\affiliation{\hkuaffil}

\author{Hoi Chun Po}
\affiliation{\mitaffil}
\affiliation{\ustaffil}

\author{Vincent Liu}
\affiliation{\mitaffil}
\affiliation{\berkeleyaffil}

\author{John~D.~Joannopoulos}
\affiliation{\mitaffil}
%
\author{Liang Fu}
\affiliation{\mitaffil}

\author{Marin~Solja\v{c}i\'{c}}
\affiliation{\mitaffil}

\date{\today} 

\begin{abstract}
    The Hofstadter model exemplifies a large class of physical systems characterized by particles hopping on a lattice immersed in a gauge field. 
    Recent advancements on various synthetic platforms have enabled highly-controllable simulations of such systems with tailored gauge fields featuring complex spatial textures.
    These synthetic gauge fields could introduce synthetic symmetries that \swap{seldom}{do not} appear in electronic materials.
    Here, in an ${\rm SU}(2)$ non-Abelian Hofstadter model, we theoretically show the emergence of multiple nonsymmorphic chiral symmetries, which combine an internal unitary anti-symmetry with fractional spatial translation. 
    Depending on the values of the gauge fields, the nonsymmorphic chiral symmetries can exhibit non-Abelian algebra and protect Kramer quartet states in the bulk band structure, creating general four-fold degeneracy at all momenta. 
    These nonsymmorphic chiral symmetries protect double Dirac semimetals at zero energy, which become gapped into quantum confined insulating phases upon introducing a boundary.
    Moreover, the parity of the system size can determine whether the resulting insulating phase is trivial or topological. 
    Our work indicates a pathway for creating topology via synthetic symmetries emergent from synthetic gauge fields. 
\end{abstract}

\maketitle


The quantum Hall~\cite{cage2012quantum} and quantum anomalous Hall \cite{haldane1988model,chang2013experimental}  effects represent the earliest examples of topological phases of matter. However, such phases with robust chiral edge modes are only realizable under stringent conditions, like a strong breaking of time-reversal symmetry, either though external magnetic fields or suitable intrinsic magnetic order. 
The topological landscape changed completely with the advent of topological insulators \cite{hasan2010colloquium,qi2011topological}. A key insight from the early studies was how time-reversal symmetry could protect new forms of nontrivial topology and this greatly enlarges the physical setups in which topological phases could emerge. Along with the particle-hole and chiral symmetries, the time-reversal symmetry represents one of the three internal symmetries relevant for the classification of topological phases, and general classification results were soon obtained under the ten-fold way \cite{kitaev2009periodic,chiu2016classification}.
The classification was then further refined in the presence of symmorphic~~\cite{fu2011topological,hsieh2012topological,tanaka2012experimental,dziawa2012topological,ando2015topological} and nonsymmorphic~\cite{fang2015new,liu2014topological,shiozaki2015z,parameswaran2013topological,varjas2015bulk,shiozaki2016topology,liang2017observation,chang2017mobius,ma2017experimental,schoop2016dirac,schoop2018tunable, wieder2016double,wieder2018wallpaper,wang2016hourglass,armitage2018weyl} spatial symmetries.
Such successive extension of the symmetry setting has led to a comprehensive understanding of the diverse set of phases protected by the 230 spatial symmetry groups~\cite{Kruthoff2017topological, po2017symmetry,bradlyn2017topological, Zhang2019catalogue, vergniory2019complete, tang2019comprehensive}, 
and the results were further extended to magnetic materials~\cite{watanabe2018structure,elcoro2021magnetic,xu2020high} in which time-reversal can also combine nontrivially with partial translation into a symmetry of the magnetic order.

Exhaustive as it may seem, the systematic treatment of (magnetic) spatial symmetries has thus far focused on symmetries that are relevant to electronic materials. Engineered physical platforms~\cite{aidelsburger2018review}, like cold-atomic, photonic, and acoustic systems, could inherently feature synthetic symmetries that would have been unnatural or fine-tuned for electronic problems~\cite{el2018non,peng2016anti,miri2013supersymmetric,zhao2020z,li2014photonic,xue2022projectively,arkinstall2017topological,kremer2020square}. 
Here we show that a non-Abelian Hofstadter model with ${\rm SU}(2)$ gauge fields, potentially realizable in engineered systems, calls for a further extension of symmetry analysis. A key new ingredient is the coexistence of multiple nonsymmorphic chiral symmetries, which combine site-dependent, local phase factors with fractional translation.
As the nonsymmorphic chiral symmetry is not generally respected in electronic materials, it is not included as part of the comprehensive (magnetic) space-group symmetry analysis on electronic topological band theory.
Yet its presence has been recognized in the study of specific models, including a minimal two-band model for nonsymmorphic topological crystalline insulators~\cite{shiozaki2015z}, certain antiferromagnetic semimetals~\cite{brzezicki2017topological}, the low-energy states in the SnTe material class~\cite{brzezicki2019topological}, and, as a theoretical construction via the square-root operation from parent Hamiltonians~\cite{arkinstall2017topological}.
So far, efforts have been mostly dedicated to systems that obey a single nonsymmorphic chiral symmetry.

In this work, we show multiple coexisting nonsymmorphic chiral symmetries could be non-Abelian, lead to intriguing symmetry algebras, and, consequently, protect unusual band degeneracy and topology, as illustrated in the non-Abelian Hofstadter model. 
More concretely, we analyze the associated algebras of the multiple nonsymmorphic chiral symmetries and reveal their dependence on the parity of the two non-Abelian gauge fields that are assumed rational.
In particular, when both of the rational gauge fields have even denominators, and one and only one of the denominators is an integer multiple of four, their algebra becomes non-Abelian and gives rise to generic Kramer quartets, \ie four-fold degeneracy at all momenta, which are jointly protected by inversion and time-reversal.
The nonsymmorphic chiral symmetries turn the magnetic Brillouin torus into a real projective plane, and therein protect double Dirac semimetals at half filling.
For relatively small systems relevant to engineered physical platforms, we further show that the semimetals get gapped and become an insulator upon the introduction of a boundary, which necessarily breaks some of the nonsymmorphic chiral symmetries.
The resulting insulator can be tuned to be either trivial or topological, depending on the parity of the system size.


\begin{figure*}[htbp]
 \includegraphics[width=\linewidth]{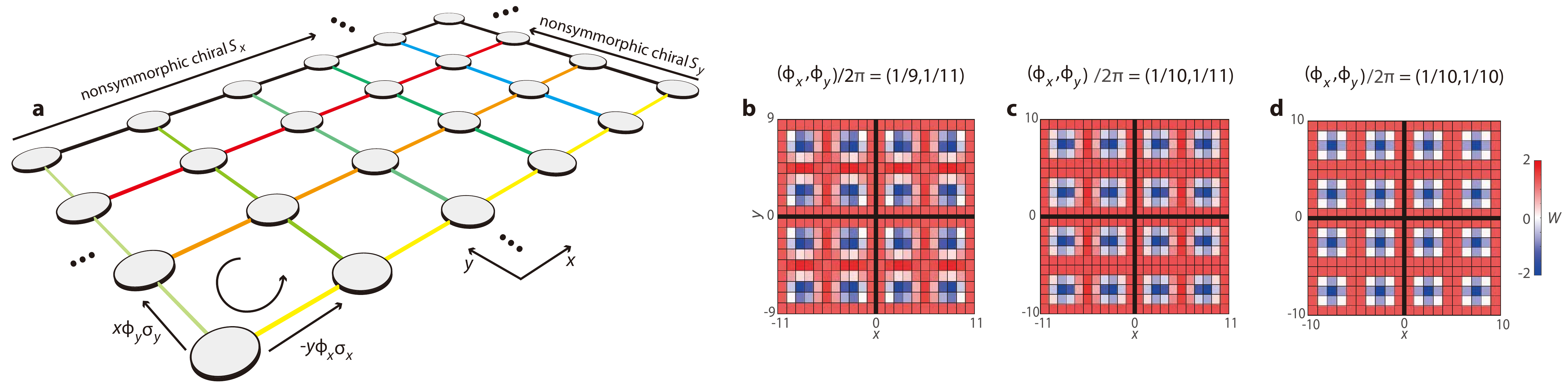}
        \caption{
        \textbf{Non-Abelian Hofstadter model (\textbf{a}) and its inhomogeneous real-space magnetic fields (\textbf{b-d}).}
        Zero, one, and two nonsymmorphic chiral symmetries involving a half-lattice translation appear in b, c and d, respectively, as evident from the patterns of the magnetic fields. Four magnetic unit cells (divided by bold black lines) are shown. 
        The real space magnetic fields are characterized by the gauge-invariant, real-space Wilson loop $W=\mathrm{Tr} \exp\left(\iu\oint\matr{A}\diff{l}\right)$ for each unit plaquette.
        }
\label{fig:magnetic_fields}
\end{figure*}

\begin{table*}
\centering
    \begin{tabular}{|c|c|c|c|}
    \hline
         $\mathbf{U}\subset\mathbf{G}$ & $q_x\in 2\mathbb{Z}+1$  & $q_x\in 4\mathbb{Z}+2$ & $q_x\in 4\mathbb{Z}$ \\
        \hline
        $q_y\in 2\mathbb{Z}+1$ & \backslashbox{}{} & $\Z_1\subset\Z_2$ & $\Z_1\subset\Z_2$ \\
        \hline
        $q_y\in 4\mathbb{Z}+2$ & $\Z_1\subset\Z_2$ & $\cellcolor{gray!30} \Z_4\times\Z_2\subset\mathcolor{blue}{\mathbb{D}_8\times\Z_2}$ & $\cellcolor{gray!30} \mathcolor{blue}{\mathbb{D}_8}\subset\mathcolor{blue}{\mathbb{D}_8\times\Z_2}$ \\
        \hline
        $q_y\in 4\mathbb{Z}$ & $\Z_1\subset\Z_2$ & $\cellcolor{gray!30} \mathcolor{blue}{\mathbb{D}_8}\subset\mathcolor{blue}{\mathbb{D}_8\times\Z_2}$ & $\cellcolor{gray!30} \mathbb{K}_4\subset\Z_2\times\Z_2\times\Z_2$ \\
        \hline
    \end{tabular}
    \quad
    \begin{tabular}{|c|c|c|c|}
    \hline
         $H$ & $q_x\in 2\mathbb{Z}+1$  & $q_x\in 4\mathbb{Z}+2$ & $q_x\in 4\mathbb{Z}$ \\
        \hline
        $q_y\in 2\mathbb{Z}+1$ & AII & DIII & CII \\
        \hline
        $q_y\in 4\mathbb{Z}+2$ & DIII & $\cellcolor{gray!30} \mathrm{AII}\oplus\mathrm{AII}\oplus\mathrm{AII}\oplus\mathrm{AII}$ & $\cellcolor{gray!30} \mathrm{CII}\oplus\mathrm{CII}$ \\
        \hline
        $q_y\in 4\mathbb{Z}$ & CII & \cellcolor{gray!30} $\mathrm{CII}\oplus\mathrm{CII}$ & $\cellcolor{gray!30} \mathrm{CII}\oplus\mathrm{CII}\oplus\mathrm{CII}\oplus\mathrm{CII}$ \\
        \hline
    \end{tabular}
    \caption{
    \textbf{Algebra of chiral symmetries and the resulting classification of the non-Abelian Hofstadter problem in the symmetric gauge.}
    $\mathbf{U}$ (left), a subgroup of $\mathbf{G}$, contains only unitary symmetries generated by the chiral symmetries $\mathbf{S}$ and results in different classifications of the systems (right). Shaded entries correspond to cases where multiple nonsymmorphic chiral symmetries appear. Blue color indicates non-Abelian groups. 
}
    \label{tab:class}
\end{table*}

\emph{Non-Abelian Hofstadter problem.}
A less-known fact about the well-known Abelian Hofstadter model, featuring U(1) gauge fields, is that it obeys chiral symmetries that are nonsymmorphic (see \sisec{sec:abelian}). 
Nevertheless, its nonsymmorphic chiral operators can always be transformed into a local basis and their algebra is always trivial (see \sisec{sec:abelian}), which relates to the equivalence between its formulations in the Landau and the symmetric gauges.
However, these conditions can get modified in non-Abelian Hofstadter models~\cite{yang20202d}.
Let us consider a Hofstader--Harper-like, SU(2) gauge fields
\begin{align}
    \matr{A} = (-y\phi_x\sigmax,x\phi_y\sigmay,0).
    \label{eq:gauge_fields}
\end{align}
The associated Hamiltonian $H(\phi_x/2\pi,\phi_y/2\pi)$ on a square lattice is given by
\begin{align}\label{eq:H_symmetric}
   H = -\sum_{x,y} t_x c^{\dagger}_{x+1,y}\e^{-\iu y\phi_x\sigmax} c_{x,y} + t_y c^{\dagger}_{x,y+1}\e^{\iu x\phi_y\sigmay}c_{x,y}+\text{H.c.}
\end{align}
Here $t_{x}$ and $t_y$ are the real hopping terms in the $x$ and $y$ directions and we restrict ourselves to $t_x=t_y=t$. $c_{x,y}$ and $c^{\dagger}_{x,y}$ are the annihilation and creation operators at site $(x,y)$. $\phi_x $ and $\phi_y$ are the SU(2) Peierls phases.
The Hamiltonian has a spin-rotation symmetry and stays invariant for arbitrary choices of distinct Pauli matrices in Eq.~\eqref{eq:gauge_fields}.
If $\phi_x$ and $\phi_y$ are both rational, \ie they can be written as
$
\phi_x = 2\pi p_x/q_x
$
and
$
    \phi_y = 2\pi p_y / q_y
$
for integers $p_x$, $q_x$, $p_y$, and $q_y$. The Hamiltonian can be solved in the $q_y \times q_x$ super-cell with the magnetic Brillouin zone (MBZ) defined as $k_x\in[0,2\pi/q_y)$ and $k_y\in[0,2\pi/q_x)$.

This model is reminiscent of but distinct from the symmetric-gauge, Abelian Hofstadter problem.
Evidently, $\hdd$ reduces to two decoupled Abelian counterparts with opposite homogeneous magnetic fluxes when either $\phi_x$ or $\phi_y$ vanishes. 
In contrast, under gauge fields that meet the genuine non-Abelian condition~\cite{yang20202d}, the associated magnetic fields become spatially inhomogeneous, as characterized by the real-space Wilson loop $W=\mathrm{Tr} \exp\left(\iu\oint\matr{A}\diff{l}\right)$ (see Fig.~\ref{fig:magnetic_fields}). 

\begin{figure}[htbp]
 \includegraphics[width=1\linewidth]{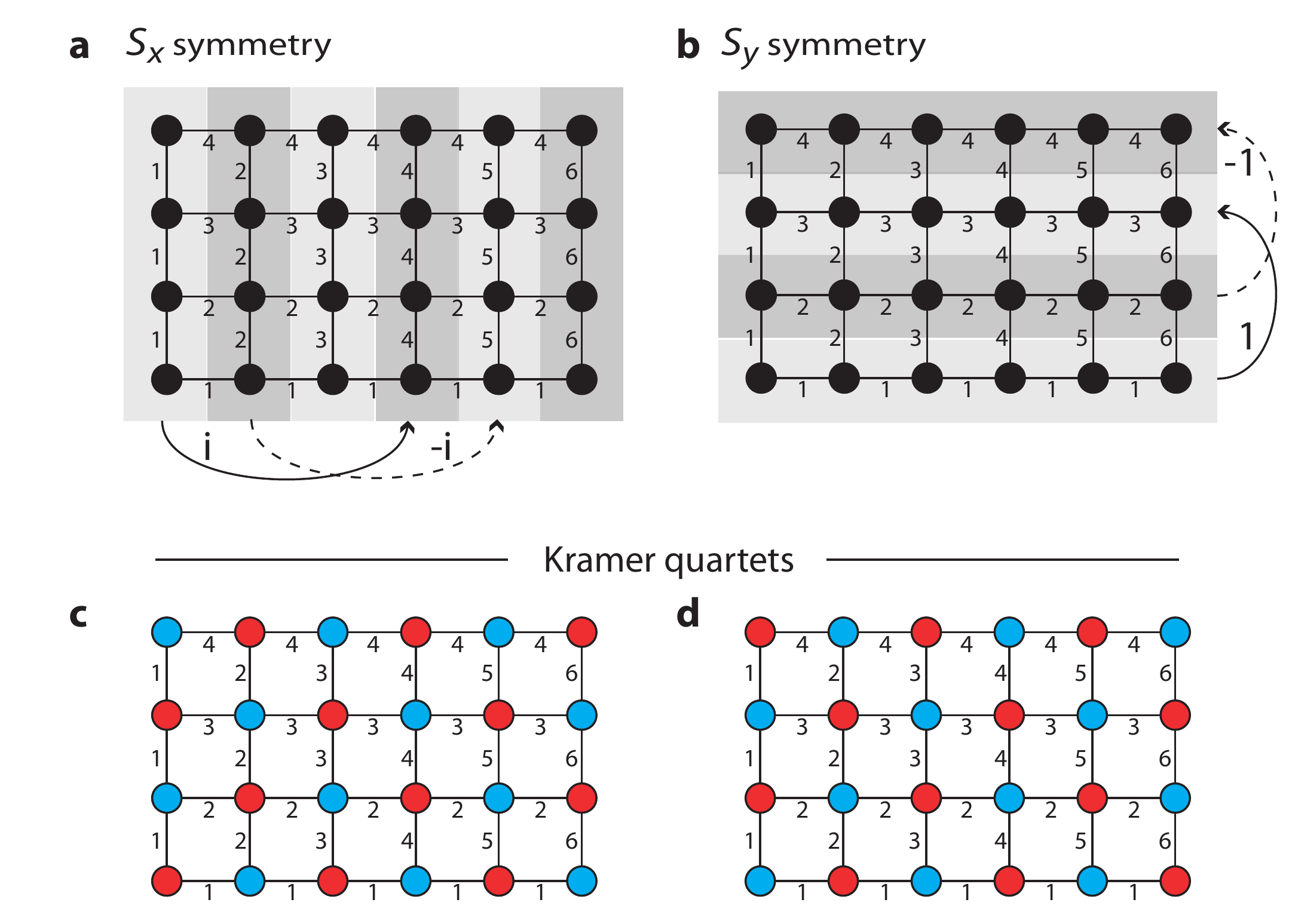}
        \caption{
        \textbf{Nonsymmorphic chiral symmetries and wavefunctions of Kramer quartets.}
        Nonsymmorphic chiral symmetries [\textbf{a} and \textbf{b}; see Eq.~\eqref{eq:ns_chiral}] appear as the product between half translations and site-dependent phase factors (represented by light and dark shadings for $\pm\iu$ and $\pm 1$ in a and b, respectively).
        When $q_x = 4\Z$ and $q_y=4\Z+2$, two Kramer doublets (\textbf{c} and \textbf{d}) can be constructed by occupying each bipartite sublattices respectively with $\alpha^{\pm}$ [ red and blue dots label the $\pm$ superscript; see Eq.~\eqref{eq:transformation_odd}] or vice versa.
        The non-Abelian group $\mathbb{D}_8$ protects the degeneracy between the two Kramer doublets and thus enables a Kramer quartet, \ie a four-fold degeneracy at all momenta and all energies.
        Here, $H\left(1/4,1/6\right)$ is illustrated as an example. 
        }
\label{fig:kramer}
\end{figure}

\emph{Symmetry algebra.}
For $\left\{\mu,\nu\right\}=\left\{x,y\right\}$, a chiral symmetry $S_\mu$ appears when $q_\nu\in 2\mathbb{Z}$. Its explicit form is given by
\begin{align}
   \bra{\mathbf{r}}S_\mu(\mathbf{k})\ket{\mathbf{r'}}  &= (-1)^{\mu}(\iu)^{q_\nu/2} \exp(\iu k_\mu q_\nu/2)\sigma_0\delta_{\mu+q_\nu/2,\mu'}\delta_{\nu,\nu'},
   \label{eq:ns_chiral}
\end{align}
where $\mathbf{r}=(x,y)$ and $\sigma_0$ operates on the spin degree of freedom.
This chiral operator contains site-dependent phase factors and a half translation along a single dimension in the magnetic unit cell (see Fig.~\ref{fig:kramer}a and b), satisfies $S_\mu^2(\mathbf{k})=\exp(\iu k_\mu q_\nu)$ that restores a full translation, and thus is an order-two nonsymmorphic symmetry with $d_\para=0$~\cite{shiozaki2016topology} for the associated Bloch Hamiltonian.
If $\left(q_x\in 2\Z, q_y \in 2\Z\right)$, multiple chiral symmetries appear. 
First, two nonsymmorphic chiral operators $S_x$ and $S_y$ emerge because we can apply Eq.~\eqref{eq:ns_chiral} to both $x$ and $y$ directions (see Fig.~\ref{fig:kramer}a and b).
\inset{It is noted that they have no counterparts in the Landau-gauge non-Abelian Hofstadter model~\cite{yang20202d}, where gauge fields are arranged along a single spatial dimension.}
Second, because the magnetic unit cell becomes bipartite under the same condition, the conventional local chiral symmetry $S_0$ exists:
\begin{align}
\bra{\mathbf{r}}S_0\ket{\mathbf{r'}}=(-1)^{x+y}\delta_{x,x'}\delta_{y,y'}\sigma_0.
\label{eq:local_chiral}
\end{align}
After we quotient away the translational part in $S_x$ and $S_y$, $\mathbf{S}\equiv\left\{S_0,S_x,S_y\right\}$ becomes the generator of a finite unitary group $\mathbf{G}$. There always exists a proper \inset{index-2} subgroup $\mathbf{U}\subset\mathbf{G}$ such that $\mathbf{U}$ only contains unitary symmetries of the hamiltonian, \ie $\mathbf{U}\equiv S_\mu S_\nu$.
\inset{$\mathbf{U}$ contains symmetries that are reminiscent of the projective translational symmetry~\cite{zhao2020z}, which are shown to protect doubly-degenerate bands and Dirac points at single momenta~\cite{zhao2020z}, as recently demonstrated in acoustic lattices~\cite{li2022acoustic,xue2022projectively}.}

The parity of the gauge field, \inset{in particular the denominators}, strongly affects the symmetry algebra and thereby the classification of this non-Abelian model~(see Table~\ref{tab:class}), even in the continuum limit with weak fields (\ie $\phi_x,\phi_y\to 0$).
If $\left(q_x\in 2\Z+1, q_y \in 2\Z+1\right)$, no other internal symmetry beside time reversal $T_0=\iu \sigma_y K$ exists, $\mathbf{U}$ is empty, and $\hdd$ belongs to class AII in Table~\ref{tab:class}.
If $\left(q_\mu=2\mathbb{Z}+1,q_\nu=2\mathbb{Z}\right)$, a nonsymmorphic chiral symmetry $S_\mu$ exists.
Although $T_0^2=-1$ and $S_\mu^2=1$ are ensured, the square of their associated nonsymmorphic particle-hole symmetry $C_{\mu}=T_0^{-1}S_{\mu}$ is uncertain---$C_\mu^2=1$ if $q_\nu=4\mathbb{Z}+2$ and $C_\mu^2=-1$ if $q_\nu=4\mathbb{Z}$, which corresponds to class DIII and CII in Table~\ref{tab:class}, respectively. For both of the DIII and CII classes, $\mathbf{G}=\{S_\mu,1\}$ is $\Z_2$ and $\mathbf{U}=\{1\}$ is the trivial group $\Z_1$. 

Richer symmetry algebra appears for $\left(q_x\in 2\Z, q_y \in 2\Z\right)$.
In this case, $T_0$ and $S_0$ enable a local particle-hole symmetry $C_0 = T_0^{-1}S_0$ that satisfies $C_0^2=-1$. 
Although $\comm{S_x}{S_y}=0$ is ensured, $S_0$ and $S_{\mu}$ may commute or anticommute.
Specifically, $\comm{S_0}{S_\mu}=0$ when $q_\nu=4\Z$ and $\acomm{S_0}{S_\mu}=0$ when $q_\nu=4\Z+2$.
There are three resulting scenarios.
First, when $\left(q_x\in 4\Z,q_y\in4\Z\right)$, $\mathbf{S}$ is an Abelian generator. $\mathbf{G}$ is an elementary Abelian group $\Z_2\times \Z_2\times \Z_2$ and $\mathbf{U}=\left\{1,S_0S_x,S_0S_y,S_xS_y\right\}$ is the Klein four-group $\mathbb{K}_4$. $H$ is consequently diagosed as $\mathrm{CII}\oplus\mathrm{CII}\oplus\mathrm{CII}\oplus\mathrm{CII}$, with each subspace chiral symmetric.
Second, when $\left(q_x\in 4\Z+2,q_y\in 4\Z+2\right)$, $\mathbf{G}=\mathbb{D}_8\times\Z_2$ becomes non-Abelian. However, its subgroup $\mathbf{U}=\Z_4\times\Z_2$ is still Abelian, which diagnoses $H$ as $\mathrm{AII}\oplus\mathrm{AII}\oplus\mathrm{AII}\oplus\mathrm{AII}$, forming two pairs of chiral partners.
Third, when $\left(q_\mu\in 4\Z+2,q_\nu\in 4\Z\right)$, $\mathbf{G}=\mathbb{D}_8\times\Z_2$ is the same as that in the previous scenario. However, they differ in their detailed symmetry algebra (\sisec{sec:sym_diff}); accordingly, the only non-Abelian subgroup of $\mathbf{U}=\mathbb{D}_8$ appears in Table~\ref{tab:class}, which leads to a classification of $\mathrm{CII}\oplus\mathrm{CII}$ and the appearance of the Kramer quartet states, as we describe below.

\begin{figure}[htbp]
 \includegraphics[width=1\linewidth]{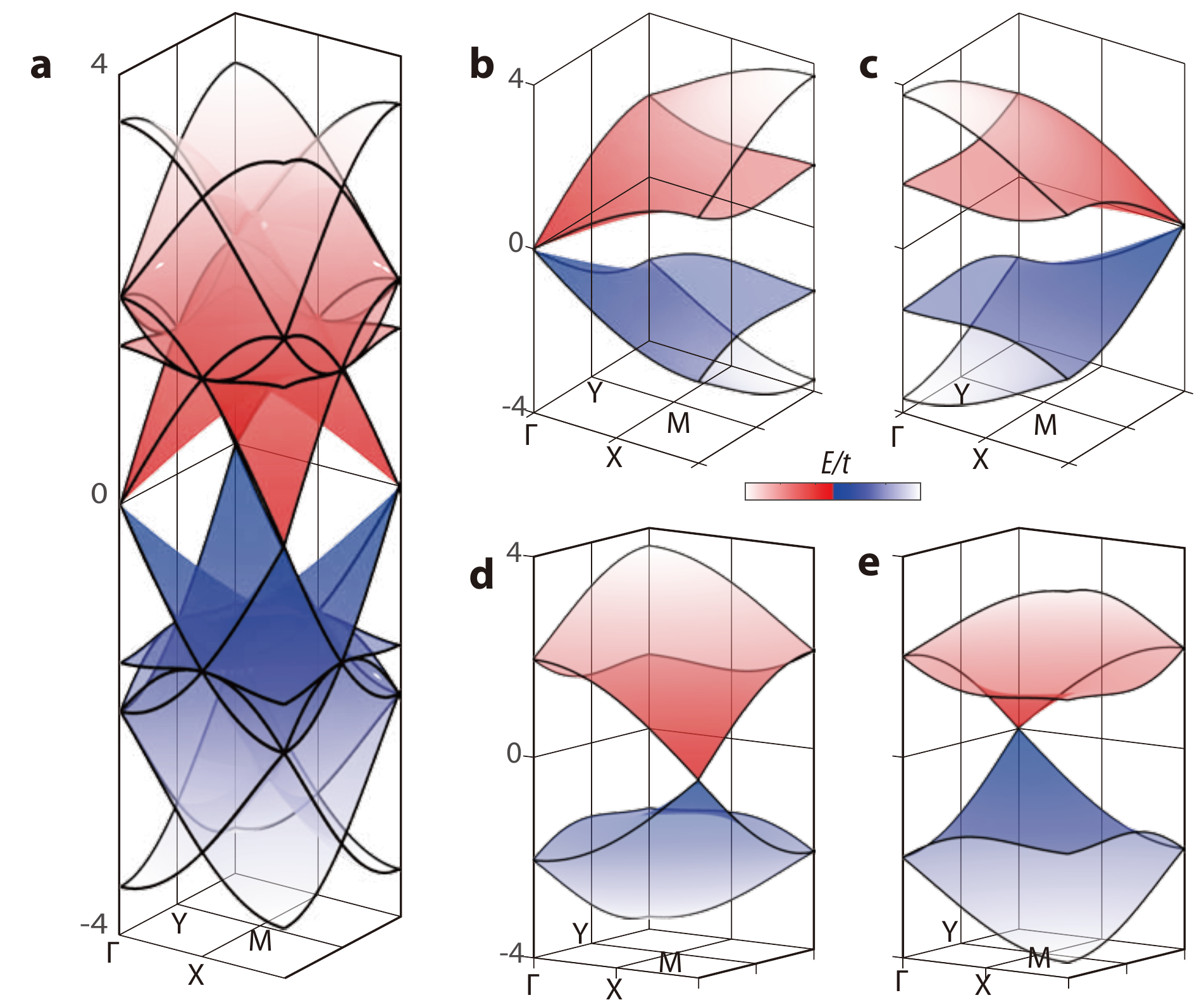}
        \caption{
        \textbf{Double Dirac points on a real projective plane protected by chiral symmetries.}
        \textbf{a.} Bulk bands and the four subspaces (\textbf{b-e}) of $\hdd(1/4,1/4)$ in the MBZ.
        All subspaces are aperiodic in the original MBZ, which becomes a real projective plane due to the nonsymmorphic symmetries along both $k_x$ and $k_y$ directions.
        The double Dirac point at $\Gamma$ of $\hdd$ is `shared' by its four subspaces at $\Gamma$, $2X$, $2Y$, and $2M$.
        }
\label{fig:bulk}
\end{figure}

\emph{Kramer quartets.}
\inset{
A consequence of the non-Abelian symmetry algebra is the existence of generic Kramer quartet states, \ie a global four-fold degeneracy in the entire MBZ for the case of $\left(q_\mu\in 4\Z+2,q_\nu\in 4\Z\right)$. Here, time reversal (can be nonsymmorphic, as constructed from the nonsymmorphic chiral symmetries) and inversion protect the conventional Kramer doublets, while the non-Abelian chiral symmetry group ensures that they are degenerate Mobius partners.}
Without loss of generality, we can decompose $H$ into two CII subspaces, $\ha$ and $\hb$, by diagonalizing the unitary symmetry $\mtp=S_xS_y$ such that both $\ha$ and $\hb$ also obey inversion and a nonsymmorphic time-reversal symmetry $T_\mu\equiv S_\mu C_{0}^{-1}$ (see~\sisec{sec:kramer}), meaning that spins at one site need to perform fractional translation to find their time-reversal partners. 
Therefore, both $\ha$ and $\hb$ are Kramer doublets.

\inset{Mathematically, the anticommutation relation in the non-Abelian group $\mathbf{U}$ enforces the degeneracy between the two CII subsystems $\ha$ and $\hb$~\cite{zhao2020z}.}
Physically, we explicitly construct their wavefunctions by considering the transformation:
\begin{align}
    \left(
    \begin{array}{c}
         \alpha^+_{mn}  \\
         \alpha^-_{mn} 
    \end{array}
    \right) = \exp(\iu\tau_x\pi/4)
    \left(
    \begin{array}{c}
         u_{mn}  \\
         u_{m+q_y/2,n+q_x/2} 
    \end{array}
    \right),    
    \label{eq:transformation_odd}
\end{align}
where $u$ is the wavefunction in the original basis, $\tau_x$ acts on the lattice sites, and we drop the spin index $\sigma$ since the transformation does not act on the spin.
$\alpha^{\pm}_{mn}$ are eigenstates of $\mtp$ with eigenvalues $\pm 1$ because
$    \mtp\alpha^{\pm}_{mn}=(-1)^{m+n}\iu^{(q_x+q_y)/2\mp 1}\alpha^{\pm}_{mn}.
    \label{Eq:eigen_odd}
$
Since $(q_x + q_y)/2=2\Z+1$, the eigenstate of $\ha$ ($\hb$) occupies each bipartite sublattice with $\alpha^{\pm}$ ($\alpha^{\mp}$), respectively (see Fig.~\ref{fig:kramer}c and d).
The two subspaces must be similar because $U_{xy}$ consists of two half translation operations, each along $x$ and $y$ directions, which map between the two sublattices (see Fig.~\ref{fig:kramer}a and b).
\inset{Taken together, the non-Abelian subgroup $\mathbf{U}$, inversion, and nonsymmorphic time-reversal symmetry jointly protect the generic Kramer quartet states at arbitrary momenta.}
\comment{ap}{I think it is good to emphasize the meaning of that again here, e.g.,, ``...protect the bulk Kramer quartet states with four-fold degeneracy at all momenta.''} 
\comment{ap}{On second thought, maybe we could contemplate on alternative terminology. My issue with ``bulk'' Kramers quartet is that, typically, when we say ``bulk'' in this context we think of the bulk of the (meta-)material in the real-space. That's very different from what you mean here, referring to the ``bulk'' of the BZ in contrast to the usual ``corners/ lines'' Kramers quartet at the boundary of the BZ. While your terminology certainly makes sense, it may not be clear for an uninitiated reader... What do you think about ``generic Kramer quartet''? That refers to ``generic momentum'' in the BZ, also being ``generic'' sounds more impressive than being ``bulk'' (to me).}

\begin{figure*}[htbp]
 \includegraphics[width=0.93\linewidth]{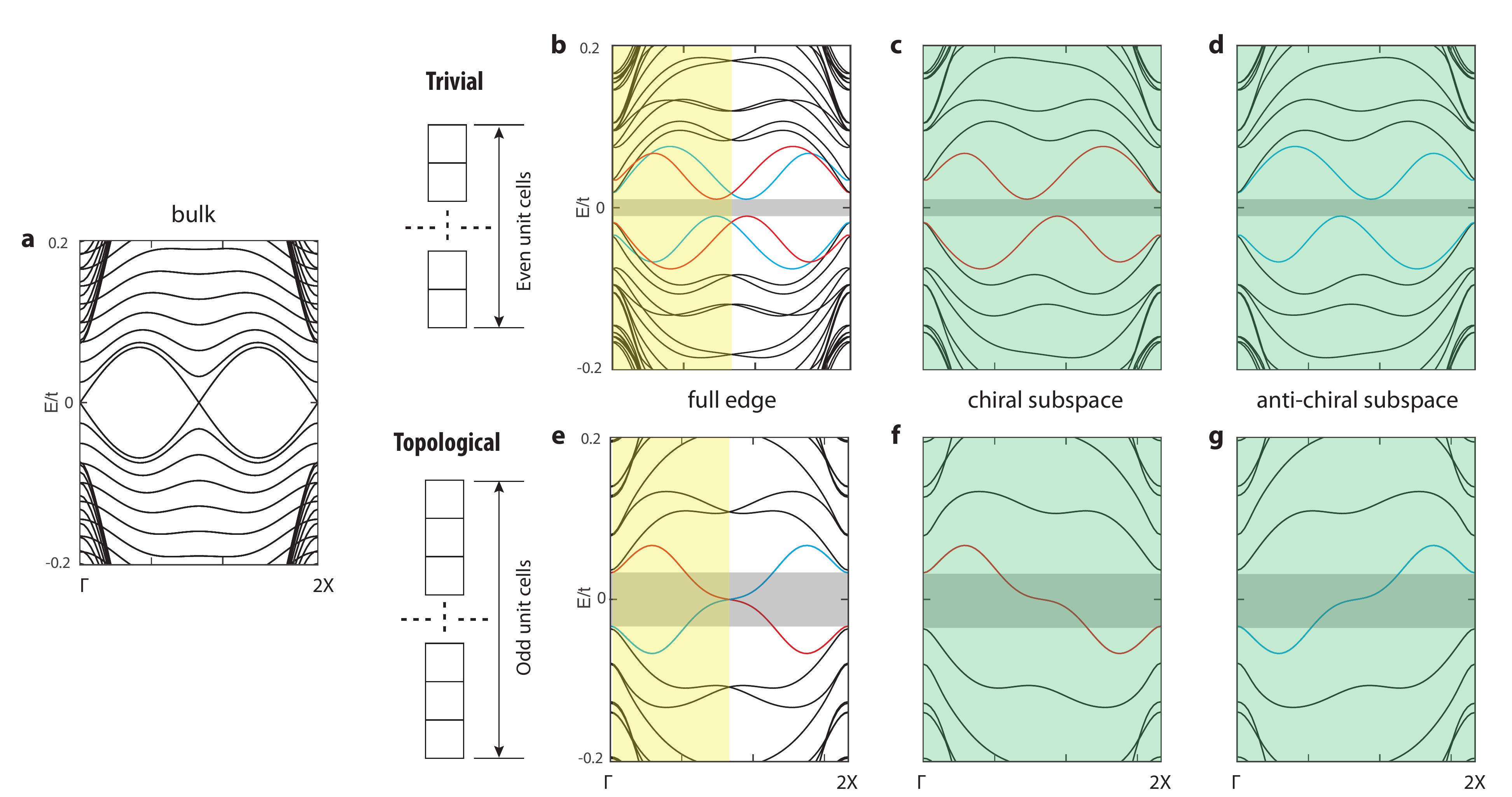}
        \caption{
        \textbf{Trivial and topological quantum-confined phases dictated by the system parity.}
        \textbf{a.} Projected bulk bands of $\hdd(1/4,1/6)$ near half-filling.
        \textbf{b-g.} quantum-confined bandgaps (shaded grey) and edge states (red and blue, for chiral and anti-chiral subspaces, respectively) appear upon introducing a boundary that breaks a nonsymmorphic chiral symmetry $S_y$, which protects the bulk double Dirac points at $\Gamma$ and $X$.
        The quantum-confined insulating phase can be trivial ($\Z_2$-even; \textbf{b-d}) and topological ($\Z_2$-odd; \textbf{e-g}) in both the full manifold and its chiral and anti-chiral subspaces when the open $y$ direction of the cylinder geometry contains an even or odd number of unit cells.
        Shaded yellow and green indicates half MBZ for the full manifold and the two subspaces.
        }
\label{fig:1416}
\end{figure*}

\emph{Quantum confinement effects on the real projective plane.}
When $\left(q_x\in 2\Z, q_y\in 2\Z\right)$ (shaded entries in Table~\ref{tab:class}), the multiple NCS symmetries $\bS$ render the MBZ of the full system a real projective plane, \ie a 2D generalization of the non-orientable Mobius strip.
For the subspaces of the system, there are two sets of momentum labels $\left[\pm \exp \left( \iu k_x q_y/2\right),\pm \exp \left(\iu k_y q_x/2\right)\right]$, independent of each other (except for the the Kramer-quartet case, as proved earlier).
As a result, the subspaces are defined on an enlarged MBZ (doubled size in both dimensions) and they exchange these momentum labels at the original MBZ boundary of the full system.

The chiral symmetries $\bS$ are also essential for stabilizing the degeneracies at half filling.
It is established that the existence of a chiral and a spatial symmetry leads to a lower bound $N_S(\mathbf{k})$ on the number of chiral zero modes at momentum $\mathbf{k}$~\cite{koshino2014topological} (also see~\sisec{sec:index}).
In our case, the spatial symmetry is inversion, which always commutes with all chiral symmetries (see \sisec{sec:kramer}).
In the presence of multiple chiral symmetries, the lower bound can be calculated for each of them, \ie $\boldsymbol{N_S}(\mathbf{k})\equiv\left(N_{S_0}(\mathbf{k}),N_{S_x}(\mathbf{k}),N_{S_y}(\mathbf{k})\right)$. As we will show below, $\boldsymbol{N_S}(\mathbf{k})$ depends on the choice of gauge fields.

In the case of $\hdd(1/4,1/4)$ (Fig.~\ref{fig:bulk}), the full system can be decomposed into four chiral-symmetric subspaces that display pairwise Mobius-type band connections along both the $x$ and $y$ directions at the zone boundary.
Notably, its double Dirac point (eight-fold degenerate; Fig.~\ref{fig:bulk}a) at $\Gamma$ is, in fact, shared by its four subspaces at the four quadrants. Each of them hosts a double Dirac point at $\Gamma$, $2X$, $2Y$, and $2M$ (Fig.~\ref{fig:bulk}b-e and also see Fig.~\ref{fig:1414}), respectively.
The chiral zero mode indices at the $\Gamma$ point are $\boldsymbol{N_S}(\Gamma)=\left(8,8,8\right)$, which indicates that inversion, and both of the local and the nonsymmorphic chiral symmetries jointly protect the double Dirac points at $\Gamma$. 
This protection is confirmed by an $x$-periodic inversion-symmetric cylinder calculation in Fig.~\ref{smfig:edge1414}, which breaks $S_y$ but preserves $S_0$ and $S_x$. 
Therein, the double Dirac point in the bulk bands reduces to a single Dirac point in the edge spectra at the cost of the reduced symmetries.

There are situations when components of $\boldsymbol{N_S}$ are not identical, which cause consequences of quantum confinement.
We again use $\hdd(1/4,1/6)$ as an example. Its chiral zero mode indices are $\boldsymbol{N_S}(\mathbf{k})=(0,0,8)$, which indicates $S_y$ being crucial in stabilizing the double Dirac points in the bulk bands (Fig.~\ref{fig:1416}a). 
Thus, an inversion-symmetric cylinder geometry, which is open in $x$ but periodic in $y$, still preserves $S_y$ and remains gapless (an example shown in Fig.~\ref{smfig:edge1416}).
In contrast, a cylinder geometry open in $y$, even if inversion-symmetric, can be gapped because the nonsymmorphic chiral symmetry $S_y$ is unavoidably broken. This is confirmed in Fig.~\ref{fig:1416} with the appearance of a band gap (shaded gray) and edge states (red and blue).
Introducing the boundary also lifts the degeneracy of the Kramer quartets because the violation of $S_y$ alters the symmetry algebra.
Specifically, $\mathbf{U}\subset\mathbf{G}$ is modified as $\Z_4\subset\mathbb{D}_8$ for this edge Hamiltonian (\cf $\mathbb{D}_8\subset\mathbb{D}_8\times\Z_2$ for the bulk in Table~\ref{tab:class}).
Evidently, bulk boundary correspondence is violated, a typical feature of topological crystalline phases but in this case originating from chiral symmetries:
the bulk phase in Fig.~\ref{fig:1416}a is a double Dirac semimetal protected by inversion and nonsymmorphic chiral symmetries, while an edge that violates $S_y$ renders it a quantum-confined insulator with a vanishingly small band gap in the thermodynamic limit.

Moreover, this quantum-confined, insulating phase can be either trivial or topological, depending on the parity of the number of unit cells in the cylinder geometry, as illustrated in Fig.~\ref{fig:1416}b-g. 
Because the cylinder still respects $S_0$ and $S_x$, we can still decompose the full manifold into two $4X$-periodic, chiral and anti-chiral subspaces that are Mobius partners to each other.
Although chiral symmetry does not hold for each subspace, there remains a sublattice symmetry that maps $E(k_x)\to-E(k_x+2X)$, typical in Hofstadter problems. 
For an even (odd) number of unit cells in the open direction, each subspace contains an even (odd) number of Kramer partners. Taken together, there must be an even (odd) number of crossings at zero energy for each subspace. Therefore, the entire manifold and its two subspaces are simultaneously $\Z_2$-even or $\Z_2$-odd depending on the parity of the unit cells along the open boundary direction. 
For other gapped phases at non-half fillings, there is no such dependence. Instead, $\mathbb{Z}_2$ time-reversal-invariant insulating phases with a Mobius structure can appear for the edge spectra (Fig.~\ref{fig:gapped1416}), where chiral-partnered edge states exchange their nonsymmorphic symmetry labels at the original MBZ boundary.

\emph{Conclusion.}
We have shown that multiple nonsymmorphic chiral symmetries naturally emerge in a non-Abelian generalization of the Hofstadter model with ${\rm SU}(2)$ gauge fields.
The nontrivial commutation relations between these symmetries lead to several topological consequences: the Kramer quartet states, semimetals at half filling, and quantum-confined insulators whose topology depends on the system parity.
Richer non-Abelian algebras and topology are expected in higher dimensions, \eg the three-dimensional non-Abelian Hofstadter model where each dimension hosts a unique nonsymmorphic chiral symmetry~\cite{liu2021three} (see~\sisec{sec:3d}).
In particular, our results showcase the diverse possibilities in which internal symmetries, classified according to the ten-fold way, could be combined with spatial symmetries in physical systems beyond electronic band theory. A particularly interesting question concerns whether similar effective symmetries could be relevant to the parton description of spin liquid candidates with an emergent ${\rm SU}(2)$ gauge field.

The proposed non-Abelian Hofstadter system could be simulated with photons and cold atoms. 
In optics, relevant candidate platforms are anisotropic or bianisotropic materials with electromagnetic duality~\cite{liu2015gauge,he2016photonic,cheng2016robust,chen2019non}, and $\mathcal{PTD}$-symmetric systems~\cite{silveirinha2017p}.
In cold atoms, the two spatially-dependent gauge potentials could be realized by existing methods---such as laser-assisted tunneling~\cite{aidelsburger2013realization,miyake2013realizing}, lattice shading~\cite{hauke2012non}, or magnetic wires with spatially-modulated currents~\cite{goldman2010realistic,anderson2013magnetically}---along the two spatial dimensions.

\subsection*{Acknowledgments}

This material is supported in part by the Air Force Office
of Scientific Research under the awards number FA9550-20-
1-0115, and FA9550-21-1-0299, as well
as in part by the US Office of Naval Research (ONR) Multidisciplinary
University Research Initiative (MURI) grant
N00014-20-1-2325 on Robust Photonic Materials with High-
Order Topological Protection. This material is also based upon
work supported in part by the U. S. Army Research Office
through the Institute for Soldier Nanotechnologies at MIT,
under Collaborative Agreement Number W911NF-18-2-0048.
Y.~Y. thanks the support from the start-up fund of the University of Hong Kong and the National Natural Science Foundation of China Excellent Young Scientists Fund (HKU 12222417).

\bibliographystyle{apsrev4-2}
\providecommand{\noopsort}[1]{}\providecommand{\singleletter}[1]{#1}%

\makeatletter\@input{xx.tex}\makeatother

\end{document}


\title{
Supplementary Materials\\
Non-Abelian nonsymmorphic chiral symmetries
    }

\author{Yi~Yang}
\email{yiyg@hku.hk}
\affiliation{\mitaffil}
\affiliation{\hkuaffil}

\author{Hoi Chun Po}
\affiliation{\mitaffil}
\affiliation{\ustaffil}

\author{Vincent Liu}
\affiliation{\mitaffil}
\affiliation{\berkeleyaffil}

\author{John~D.~Joannopoulos}
\affiliation{\mitaffil}
%
\author{Liang Fu}
\affiliation{\mitaffil}

\author{Marin~Solja\v{c}i\'{c}}
\affiliation{\mitaffil}

\maketitle

\setlength{\parindent}{0em}
\setlength{\parskip}{.5em}

\noindent{\small\textbf{\textsf{CONTENTS}}}\\ 
\twocolumngrid
\begingroup 
    \let\bfseries\relax 
    \deactivateaddvspace 
    \deactivatetocsubsections 
    \tableofcontents
\endgroup
\onecolumngrid



\section{Chiral symmetries in symmetric-gauge Abelian Hofstadter problem}
\label{sec:abelian}

The symmetry algebras of the chiral symmetries in the Abelian and non-Abelian Hofstadter problems are related but different.
The Abelian Hofstadter problem on a square lattice can be solved in the symmetric gauge $\mathbf{A}=\phi\left(-y,x,0\right)$, which describes a homogeneous magnetic field of $2\phi$ (we drop the conventional factor of $1/2$ in its vector potential to maintain consistency with the non-Abelian model described in the main text). 
When $q$ is odd, all internal symmetries are broken and the Abelian Hofstadter model belongs to class A.
When $q$ is even, The model obeys chiral symmetries~\cite{wen1989winding} similar to those in the main text. 
A nonsymmorphic chiral operator $S_\mu$ written in real space is given explicitly by
\begin{align}
    \bra{\mathbf{r}}S_\mu\ket{\mathbf{r'}} = (-1)^{\mu}(\iu)^{q_\nu/2}\delta_{\mu+q_\nu/2,\mu'}\delta_{\nu,\nu'}, 
    \label{eq:abelian_chiral}
\end{align}
where $\mathbf{r}=(x,y)$, $\left\{\mu,\nu\right\}=\left\{x,y\right\}$ and $q_x=q_y=q$ in the symmetric gauge.
Meanwhile, the conventional local chiral symmetry 
\begin{align}
\bra{\mathbf{r}}S_0\ket{\mathbf{r'}}=(-1)^{x+y}\delta_{x,x'}\delta_{y,y'}
\label{eq:local_chiral}
\end{align}
also appears since the magnetic unit cell is bipartite for $q$ being even.  $\mathbf{S}\equiv\left\{S_0,S_x,S_y\right\}$ again becomes the generator of a finite unitary group $\mathbf{G}$ and its index-2 subgroup $\mathbf{U}\subset\mathbf{G}$ such that $\mathbf{U}$ only contains unitary symmetries of the Hamiltonian. Their associated algebra and classification are contained in Table~\ref{tab:abelian_class}.

\begin{table*}[htbp]
    \centering
    \begin{tabular}{|c|c|c|c|}
    \hline
        $\mathbf{U}\subset\mathbf{G}$ & $q\in 2\mathbb{Z}+1$ & $q\in 4\mathbb{Z}+2$  & $q\in 4\mathbb{Z}$  \\
        \hline
        $q\in 2\mathbb{Z}+1$ & \backslashbox{}{} & \backslashbox{}{} & \backslashbox{}{} \\
        \hline
        $q\in 4\mathbb{Z}+2$ & \backslashbox{}{} & $\Z_4\times\Z_2\subset\mathbb{D}_8\times\Z_2$ & \backslashbox{}{} \\
        \hline
        $q\in 4\mathbb{Z}$  & \backslashbox{}{} & \backslashbox{}{} & $\mathbb{K}_4\subset\Z_2\times\Z_2\times\Z_2$  \\
        \hline
    \end{tabular}
    \quad
    \begin{tabular}{|c|c|c|c|}
    \hline
       $H$  & $q\in 2\mathbb{Z}+1$ & $q\in 4\mathbb{Z}+2$  & $q\in 4\mathbb{Z}$  \\
        \hline
        $q\in 2\mathbb{Z}+1$ & $\mathrm{A}$ & \backslashbox{}{} & \backslashbox{}{}  \\
        \hline
        $q\in 4\mathbb{Z}+2$ & \backslashbox{}{} & $\mathrm{A}\oplus\mathrm{A}\oplus\mathrm{A}\oplus\mathrm{A}$ & \backslashbox{}{}  \\
        \hline
        $q\in 4\mathbb{Z}$ & \backslashbox{}{}  & \backslashbox{}{} & $\mathrm{AIII}\oplus\mathrm{AIII}\oplus\mathrm{AIII}\oplus\mathrm{AIII}$ \\
        \hline
    \end{tabular}
        \caption{
    \textbf{Algebra of chiral symmetries and the resulting classification of the Abelian Hofstadter problem in the symmetric gauge.}
    $\mathbf{U}$ (left), a subgroup of $\mathbf{G}$, contains only unitary symmetries generated by the chiral symmetries $\mathbf{S}$ and results in different classification of the systems (right).
     $\mathbf{U}$ belongs to different groups, $\mathbb{Z}_4\times\mathbb{Z}_2$ and $\mathbb{K}_4$, for $q$ being $4\mathbb{Z}+2$ or $4\mathbb{Z}$, respectively (top). Both of the two groups are Abelian and thus enable block diagonalization of the symmetric-gauge problem into a direct sum of four Landau-gauge solutions (bottom) with an extra $q$-fold degeneracy due to zone folding. 
}
    \label{tab:abelian_class}
\end{table*}
\section{Difference between the symmetry algebras of $\left(q_x\in 4\Z+2,q_y\in 4\Z+2\right)$ and $\left(q_x\in 4\Z,q_y\in 4\Z+2\right)$}\label{sec:sym_diff}
Although the chiral symmetries $\mathbf{S}$ of the two cases give the same symmetry group $\mathbf{G}=\mathbb{D}_8\times\Z_2$, they differ in the detailed symmetry algebra. 
A presentation of the $\mathbb{D}_8\times\Z_2$ group is
\begin{align}
    \mathbb{D}_8\times\Z_2\equiv\bra{a,b,c}\ket{a^4=b^2=c^2=e,bab=a^{-1},ac=ca,bc=cb}.
\end{align}
In the $\left(q_x\in 4\Z+2,q_y\in 4\Z+2\right)$ case, $a=S_0S_x$, $b=S_x$, and $c=S_x S_y$.

In the $\left(q_x\in 4\Z,q_y\in 4\Z+2\right)$ case, $a=S_0S_x$, $b=S_x$, and $c=S_y$. Note that $\comm{S_0}{S_y}=0$ and $\acomm{S_0}{S_x}=0$.

\section{Remarks on Kramer quartets}
\label{sec:kramer}

\subsection{Commutation relation of $\mathbf{S}$ with $P$ and $T_0$ symmetries}

$\mathbf{S}$, $P$, and $T_0$ operate on the wave function $u_{x,y}$ as
\begin{align}
    S_0u_{x,y}&=(-1)^{x+y}\sigma_0 u_{x,y},\\
    S_\mu u_{\mu,\nu} &=(-1)^\mu\iu^{q_\nu/2}\sigma_0 u_{\mu+q_\nu/2,\nu},\\
    T_0u_{x,y}&=\iu\sigmay u^*_{x,y},\\
    P u_{x,y} &= \sigma_0 u_{q_y-x,q_x-y}.
\end{align}
We thus have the commutation relation
\begin{align}
    \comm{T_0}{S_0}&=0,\\
    \comm{T_0}{S_\mu}u_{\mu,\nu}&=(-1)^\mu\left[(-\iu)^{q_\nu/2}-(+\iu)^{q_\nu/2}\right]\iu\sigmay K u_{\mu+q_\nu/2,\nu},\\
    \comm{P}{S_0}u_{x,y}&=\left[(-1)^{x+y}-(-1)^{(q_x+q_y)-(x+y)}\right]\sigma_0u_{q_y-x,q_x-y}=0,\\    
    \comm{P}{S_\mu}u_{\mu,\nu}&=\left[(-1)^{\mu}-(-1)^{(q_\nu-\mu)}\right]\sigma_0u_{q_\nu/2-\mu,q_\mu-\nu}=0,\\
    \comm{P}{T_0}&=0,
\end{align}
where $\left\{\mu,\nu\right\}=\left\{x,y\right\}$.

\subsection{Time-reversal and inversion symmetry for each subspace when $q_x\in 4\mathbb{Z}$ and $q_y\in 4\mathbb{Z}+2$}\label{sec:parity-time}

Without loss of generality, we consider the block diagonalization of $H$ with $\mtp$.
$H$ obeys inversion symmetry $P$ with four inversion centers for $H$ given by  $\left(\left\{0,q_y/2\right\},\left\{0,q_x/2\right\}\right)$. 
$\mtp$ and $P$ operate on the wave function as
\begin{align}
    \mtp u_{x,y} = (-1)^{x+y}\iu^{\left(q_x+q_y\right)/2}\sigma_0 u_{x+q_y/2,y+q_x/2}
\end{align}
    and
\begin{align}
    P u_{x,y} = u_{q_y-x,q_x-y}.
\end{align}
Therefore, 
\begin{align}
    \left[P,\mtp\right]=\left[(-1)^{q_y+q_x-x-y}-(-1)^{x+y}\right]\iu^{(q_x+q_y)/2} u_{q_y/2-x,q_x/2-y},
\end{align}
which vanishes for $q_x+q_y=2\Z$.
Although $\left[\mtp,T_0\right]\neq0$ for $(q_x +  q_y)/2=2\Z+1$, we can instead consider a nonsymmorphic time reversal symmetry $T_{\mu}\equiv S_{\mu}P_0^{-1}$ such that $\left[\mtp,T_{\mu}\right]=0$.
Taken together, we can simultaneously block diagonalize $H$, $T_{\mu}$, and $P$ with $V_{xy}$ such that $V_{xy}\mtp V^{\dagger}_{xy}=\mathrm{diag}(-I,I)$ and $V_{xy}H V_{xy}^\dagger = \mathrm{diag}\left(\ha,\hb\right)$, with each block obeying time-reversal and inversion symmetries simultaneously.
Therefore, $\ha$ and $\hb$ are both Kramer partnered in the entire MBZ.

\subsection{Appearance of Kramer quartet in another class}\label{sec:kramer_antiexample}

In the main text, we show that Kramer quartets must appear when $q_x=4\mathbb{Z}$ and $q_y=4\mathbb{Z}+2$.
However, the reverse statement is not generally true because Kramer quartets could also appear in $H(1/4,1/8)$ under class $\mathrm{CII}\oplus\mathrm{CII}\oplus\mathrm{CII}\oplus\mathrm{CII}$.
For this class, we need to consider another transformation
\begin{align}
    \left(
    \begin{array}{c}
         \beta^{+}_{mn}  \\
         \beta^{-}_{mn} 
    \end{array}
    \right) = \exp(\iu\tau_y\pi/4)
    \left(
    \begin{array}{c}
         u_{mn}  \\
         u_{m+q_y/2,n+q_x/2} 
    \end{array}
    \right).
    \label{eq:transformation_even}
\end{align}
Again, $\beta^{\pm}_{mn}$ are eigenstates of $\mtp$ because
\begin{align}
    \mtp\beta^{\pm}_{mn}=\mp (-1)^{x+y}\iu^{(q_x+q_y)/2}\beta^{\pm}_{mn}.
\end{align}
In this case, the appearance of four-fold, full-MBZ degeneracy becomes equivalent to the requirement that there exists a translation $(\Delta m,\Delta n)$ such that $\beta^{+}_{mn}=\beta^{-}_{m+\Delta m, n+\Delta n}$ for all $(x,y)$.
We find that this requirement is accidentally met in $\left(\phi_x,\phi_y)/2\pi=(1/4,1/8\right)$, but it is not satisfield for other choices of gauge fields in this class.

\section{Topological index for chiral zero modes}
\label{sec:index}

We adopt the topological index derived in Ref.~\cite{koshino2014topological} for the chiral zero modes in our system.
If there exists a spatial symmetry (usually inversion $P$ in this case) that satisfies
$\left[H(\mathbf{k}),P\right]=0$ and $\left[S,P\right]=0$, we can block diagonalize $H(\mathbf{k})=\mathrm{diag}\left[H_i(\mathbf{k})\right]$ and $S=\mathrm{diag}(S_i)$ (where $S$ is certain element of $\bS$) in a basis such that $P$ is diagonal and each subspace $H_i(\mathbf{k})$ respects chiral symmetry.
The number of Dirac zero modes therefore satisfies $N(\mathbf{k})\geq N_S(\mathbf{k}) \equiv \sum_i \mathrm{Tr} \, S_i$.

\section{Additional spectra}\label{sec:spectra}

\begin{figure}[htbp]
 \includegraphics[width=\linewidth]{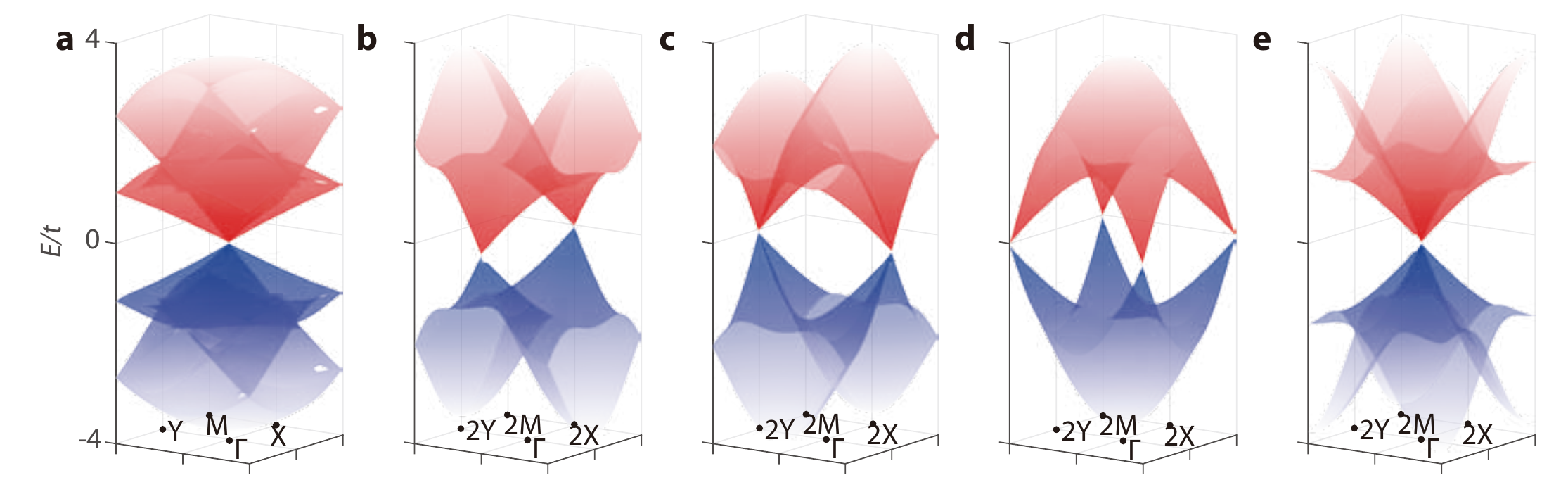}
        \caption{
        \textbf{Bulk bands in the MBZ (a) and its four subspaces (\textbf{b-e}) of $H(1/4,1/4)$ in the double-sized MBZ.}
        All subspaces are aperiodic in the original MBZ, which becomes a real projective plane because of the nonsymmorphic symmetries along both $k_x$ and $k_y$ directions.
        The double Dirac point at $\Gamma$ of $H$ is 'shared' by its four subspaces at $\Gamma$,$2X$, $2Y$, and $2M$.
        b and c are $C_4$ partners, while d and e are each $C_4$ symmetric.
        }
\label{fig:1414}
\end{figure}


\begin{figure}[htbp]
 \includegraphics[width=0.97\linewidth]{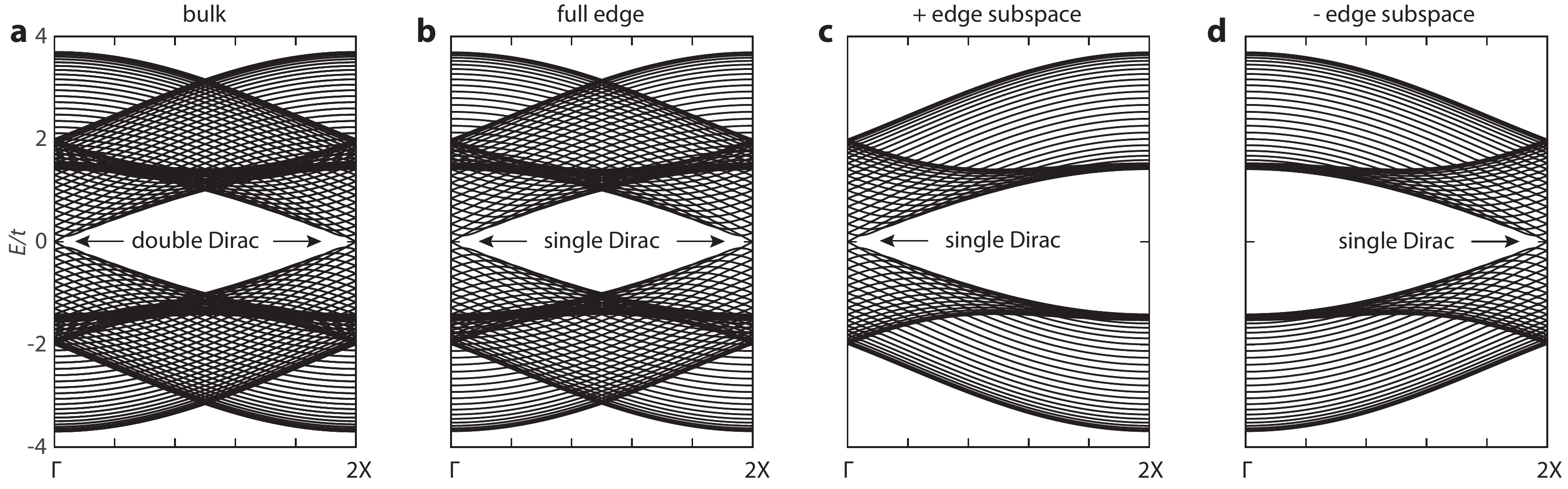}
        \caption{
        \textbf{Edge spectra for a $x$-periodic, inversion-symmetric cylinder of $H(1/4,1/4)$.}
        Projected bulk bands host a double Dirac point (\textbf{a}) at $\Gamma$, which reduces to two single Dirac points in the edge spectra, each hosted by the two edge subspaces [\textbf{c-d}; labelled by eigenvalues of $U(k_x)=S_0S_x(k_x)$].
        %
        Breaking inversion will remove the Dirac points and open a gap.
        }
\label{smfig:edge1414}
\end{figure}


\begin{figure}[htbp]
 \includegraphics[width=0.97\linewidth]{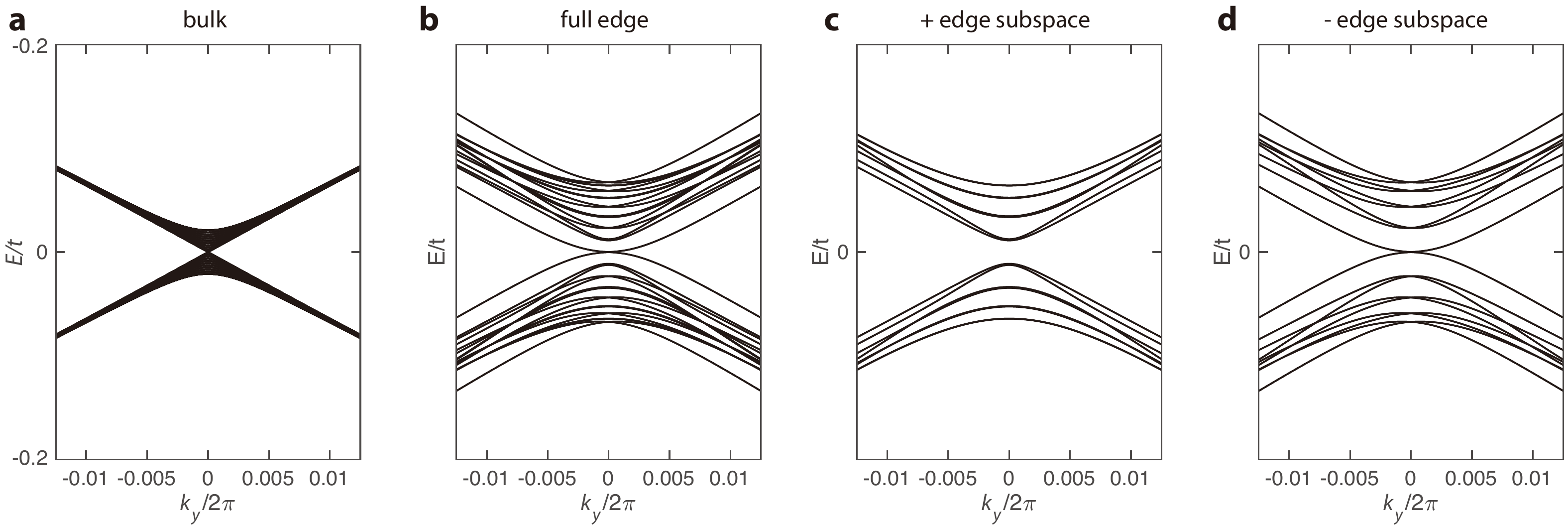}
        \caption{
        \textbf{Edge spectra for a $y$-periodic, inversion-symmetric cylinder of $H(1/4,1/6)$.}
        Projected bulk bands host a double Dirac point (\textbf{a}) at $\Gamma$, which reduces to a single quadratic Dirac point in the edge spectra, hosted by one (\textbf{d}) of the two edge subspaces [labelled by eigenvalues of $U(k_x)=S_0S_x(k_x)$]. The remaining subspace becomes gapped (\textbf{c}) at $\Gamma$, but hosts the same quadratic Dirac point at momentum $2Y$.
        }
\label{smfig:edge1416}
\end{figure}


\begin{figure}[H]
 \includegraphics[width=0.5\linewidth]{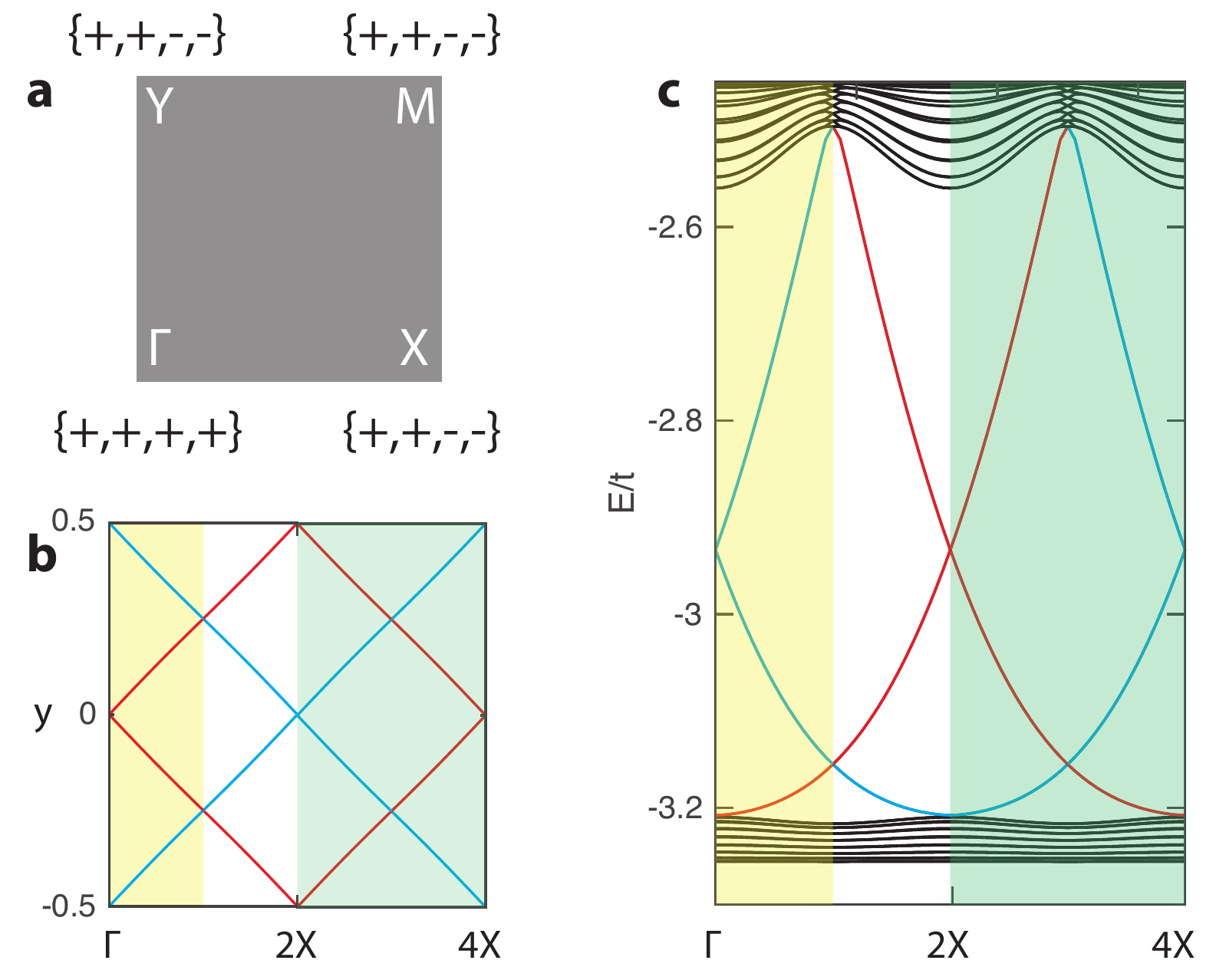}
  \centering
        \caption{
        \textbf{$\Z_2$ topological insulators with alternating nonsymmorphic symmetry labels.}
        \textbf{a.} parity eigenvalues for a filled Kramer quartet (two filled Kramer pairs, each being the Mobius partner of the other) below the lowest gap of  for $H(1/4,1/6)$. These ensemble parity eigenvalues cannot be specified for each Kramer pairs because they exchange their nonsymmorphic symmetry labels at the MBZ boundary.
        \textbf{b-c.} Bulk Wannier spectra (b) and edge spectra (c) from a cylinder geometry. 
        To diagnose the phases of the two subspaces, we calculate their Wannier spectra, as shown in Fig.~\ref{fig:gapped1416}b. Both the two subspaces, whose Wannier spectra are labelled by red and blue, respectively, are periodic in the enlarged MBZ (whose half MBZ in shaded green) and are $\Z_2$-odd. Meanwhile, their combined full Hamiltonian is also $\Z_2$-odd in the original MBZ (whose half MBZ in shaded yellow) and the two Kramer pairs switch their nonsymmorphic symmetry labels at $X$ point. This diagnosis is corroborated by the spectra of a cylinder geometry (Fig.~\ref{fig:gapped1416}c), which also indicate a $\Z_2$-odd phase for the total manifold and the two subspaces. 
         }
\label{fig:gapped1416}
\end{figure}

\section{3D model}\label{sec:3d}

Richer symmetry algebras are expected to appear for a three-dimensional (3D) non-Abelian Hofstadter model~\cite{liu2021three} featured by an inhomogeneous, Hofstadter--Harper-like, non-Abelian gauge potentials in 3D
\begin{align} \label{eq:3d_gauge}
     \mathbf{A}_{\pm} &= \left[ \left(z \pm y \right) \theta_{x} \sigma_x, \left( x \pm z \right) \theta_{y} \sigma_y, \left(y \pm x \right) \theta_{z} \sigma_z \right].
\end{align}
The associated Hamiltonian is given by
\begin{align} \label{eq:hh_ham}
    \begin{split}
        H_{\pm}(\theta_{x},\theta_{y},\theta_{z}) = - \sum_{x,y,z} & t_x c^\dagger_{x+1,y,z} e^{\iu \left(z \pm y \right) \theta_{x} \sigma_x} c_{x,y,z} 
        + t_y c^\dagger_{x, y+1, z} e^{\iu \left( x \pm z \right) \theta_{y} \sigma_y} c_{x,y,z} 
        + t_z c^\dagger_{x, y, z+1} e^{\iu \left(y \pm x \right) \theta_{z} \sigma_z} c_{x,y,z} + \mathrm{H.c..} 
    \end{split}
\end{align}
The choice of $\pm$ corresponds to different magnetic fields. Every one of its 2D Cartesian surfaces (along the $xy$, $zx$, and $yz$ planes) reduces to the 2D non-Abelian Hofstadter model studied in the main text.
This three-dimensional non-Abelian system can be solved in an enlarged magnetic unit cell if $\theta_{x}$, $\theta_{y}$, and $\theta_{z}$ are rational multiples of $2 \pi$, \ie $\theta_{x} = 2 \pi p_{x} / q_{x}$, $\theta_{y} = 2 \pi p_{y} / q_{y}$, and $\theta_{z} = 2 \pi p_{z} / q_{z}$ (where $p_i$ and $q_i$ are co-prime). The size of the magnetic unit cell is $\mathrm{lcm} \left( q_{y}, q_{z} \right) \times \mathrm{lcm} \left( q_{x}, q_{z} \right) \times \mathrm{lcm} \left( q_{x}, q_{y} \right) \equiv Q_{x} \times Q_{y} \times Q_{z}$, where $\mathrm{lcm}$ denotes the least common multiple. Consequently, the associated MBZ is $k_x \in \left[ 0, 2 \pi / Q_{x} \right)$, $k_y \in \left[ 0, 2 \pi / Q_{y} \right)$, $k_z \in \left[ 0, 2 \pi / Q_{z} \right)$.

$H_\pm$ obeys a nonsymmorphic chiral symmetry when at least one of $q_x$, $q_y$, and $q_z$ is even. Without loss of generality, we assume $Q_{x} = \mathrm{lcm} \left( q_{y}, q_{z} \right)$ is even. The associated nonsymmorphic chiral operator $S_{x}$ is
\begin{align} \label{smeq:3d_chiral}
  \bra{\mathbf{r}}S_x\ket{\mathbf{r'}} = \left( -1 \right)^{x + y \alpha_{xy} + z \alpha_{xz}} \left( \iu \right)^{Q_{x} / 2} \sigma_0 \delta_{x+Q_x/2,x'}\delta_{y,y'}\delta_{z,z'},
\end{align}
where $\alpha_{\mu\nu} \equiv \left( Q_\mu p_\nu / q_\nu + 1 \right) \mod 2$. Eq.~\eqref{smeq:3d_chiral} can be applied to all three directions to construct multiple chiral symmetries, which should enable richer chiral symmetry algebras. 


\bibliographystyle{apsrev4-1}
\bibliography{sources.bib}